\documentclass[11pt]{article}
\usepackage{graphicx}
\usepackage{epstopdf}
\def\bm{\boldmath}
\def\br{\mbox{\bm $r$}}
\def\bp{\mbox{\bm $p$}}
\def\~{\tilde}

\title{Delayed Choice Experiments and the Bohm Approach.}
\author{B. J. Hiley and R. E. Callaghan.} \date{Theoretical Physics Research Unit\\  Birkbeck
College, University of London\\ Malet Street, London WC1E 7HX,  England\footnote{b.hiley@bbk.ac.uk}
\vspace{0.5cm}}
\begin{document}
\maketitle
\begin{abstract}
The delayed choice experiments of the type introduced by Wheeler and extended by Englert, Scully, S\"{u}ssmann and Walther [ESSW], and  others, have formed a rich area for investigating the puzzling behaviour of particles undergoing quantum interference.  The surprise provided by the original delayed choice experiment, led Wheeler to the conclusion that  ``no phenomenon is a phenomenon until it is an observed phenomenon", a radical explanation which implied that ``the past has no existence except as it is recorded in the present".  However Bohm, Dewdney and Hiley have shown that the Bohm interpretation gives a straightforward account of the behaviour of the particle without resorting to such a radical explanation.  The subsequent modifications of this experiment led both Aharonov and Vaidman and [ESSW] to conclude that the resulting Bohm-type trajectories in these new situations produce unacceptable properties. For example, if a cavity is placed in one arm of the interferometer, it will be excited by a particle travelling down the {\em other}  arm.  In other words it is the particle that does {\em not} go through the cavity that gives up its energy! If this analysis is correct, this behaviour would be truly bizarre and could only be explained by an extreme non-local transfer of energy that is even stronger than that required in an EPR-type processes.  In this paper we show that this conclusion is not correct and that if the Bohm interpretation is used correctly, it gives a {\em local} explanation, which actually corresponds exactly to the standard quantum mechanics explanation offered by Englert, Scully, S\"{u}ssmann and Walther [ESSW]. 
\end{abstract}

\section{ Introduction.}

The idea of a delayed choice experiment was first introduced by Wheeler \cite{w78}  and discussed in further detail by Miller and Wheeler \cite{mw83}.  They wanted to highlight the puzzling behaviour of a single particle\footnote{We will begin the discussion using electron interference as it can be described by the Schr\"{o}dinger equation.  A discussion of photons requires field theory.  We will discuss the role of photons later in the paper.}  in an interferometer when an adjustment is made to the interferometer by inserting (or removing) a beam splitter at the last minute.  Wheeler argued that this presents a conceptual problem even when discussed in terms of standard quantum mechanics (SQM) because the results seemed to imply that there was a change in behaviour from wave-like phenomenon to particle-like phenomenon or vice-versa well after the particle entered the interferometer. 

The example Miller and Wheeler \cite{mw83} chose to illustrate this effect was the Mach-Zender interferometer shown in Figure 1. In this set up a movable beam splitter $ BS_{2}$ can either be inserted or removed just before the electron is due to reach the region $ I_{2}$. When $BS_{2}$ is not in position, the electron behaves like a particle following one of the paths, 50\% of the time triggering $D_{1}$ and the other 50\% of the time triggering $D_{2}$.  However when the beam splitter is in place, the electron behaves like a wave following ``both" paths, and the resulting interference directs all the particles into $D_{1}$.  Wheeler's claim is that by 
delaying the choice for fixing the position of the final beam splitter forces the electron to somehow ÔdecideÕ whether to behave like a particle or a wave long after it has passed the first beam splitter $BS_{1}$, but before it has reached $I_{2}$.  Experiments of this type, which have been reviewed in Greenstein and Zajonc \cite{gz97}, confirm the predictions of quantum theory and raises the question ``How is this possible?''
\begin{figure}
\begin{center}
\includegraphics[width=4in]{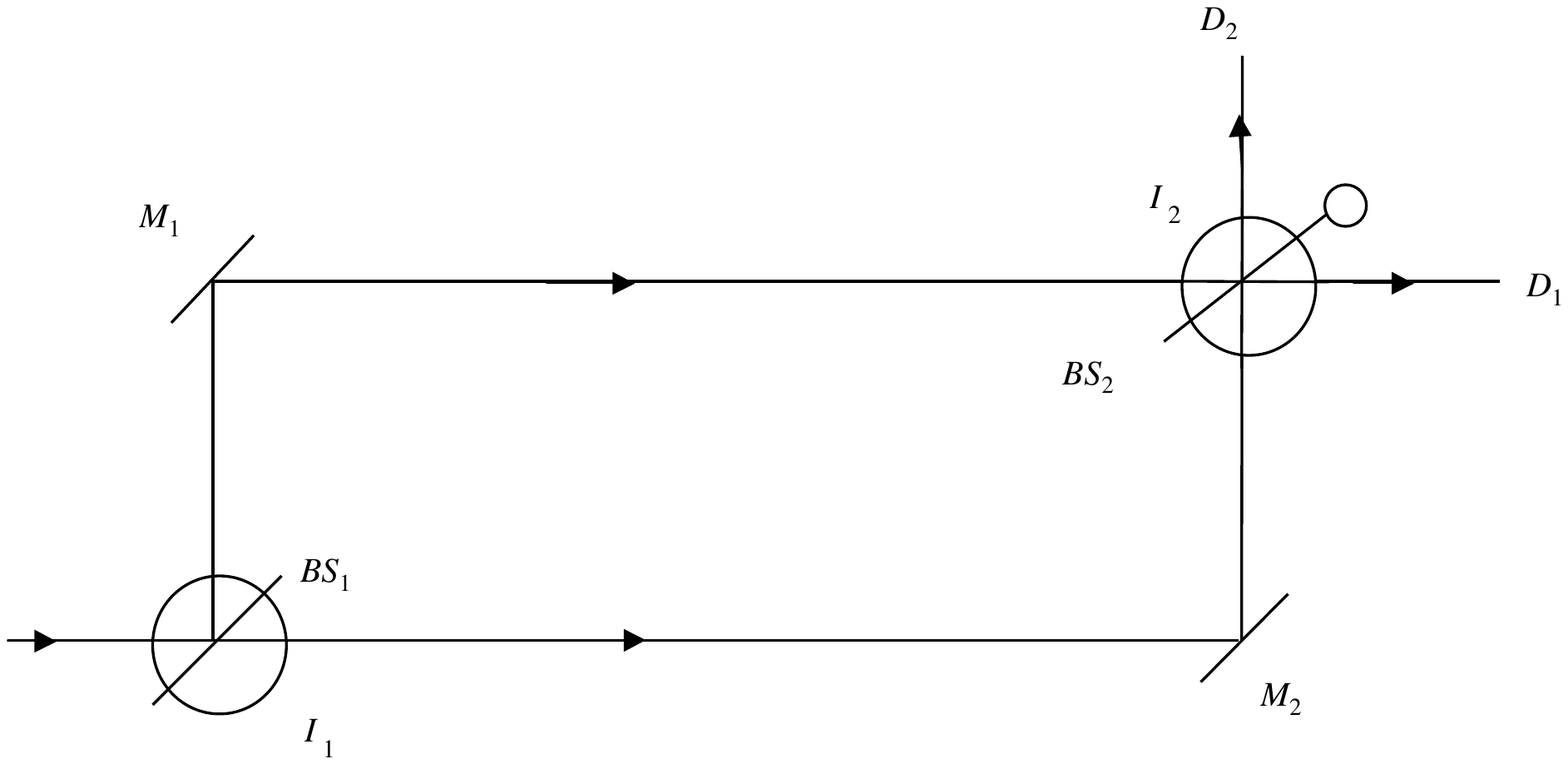}
\vspace{-4in}
\end{center}
\caption{ Sketch of the Wheeler delayed choice experiment}
\end{figure}
Wheeler \cite{w78} resolves the problem in the following way.  
\begin {quote} Does this result mean that present choice influences past dynamics, in contravention of every formulation of causality? Or does it mean, calculate pedantically and don't ask questions?  Neither; the lesson presents itself rather like this, that the past has no existence except as it is recorded in the present. \end{quote}

Although Wheeler claims to be supporting Bohr's position, Bohr \cite{b61}actually comes to a different conclusion and writes
\begin {quote} In any attempt of a pictorial representation of the behaviour of the photon we would, thus, meet with the difficulty: to be obliged to say, on the one hand, the photon always chooses {\em one} of the two ways and, on the other hand, that it behaves as if it passed {\em both} ways.\end {quote}

Bohr's conclusion is not that the past has no existence until a measurement is made, but rather that it was no longer possible to give `pictures' of quantum phenomena as we do in classical physics.  For Bohr the reason lay in the `indivisibility of the quantum of action' as he put it, which implies it is not possible to make a sharp separation between the properties of the observed system and the observing apparatus.  Thus it is meaningless to talk about the path taken by the particle and in consequence we should simply give up attempts to visualise the process.  Thus Bohr's position was, to put it crudely, `calculate because you cannot get answers to such questions', a position that Wheeler rejects. 

But it should be quite clear from the literature that many physicists even today do not accept either Bohr's or Wheeler's position and continue to search for other possibilities of finding some form of image to provide a physical understanding of what could be actually going on in these situations, hence the continuing debate about the delayed choice experiment.

By now it is surely well known that the Bohm interpretation (BI)\footnote {In this paper we will use the term `Bohm interpretation' to stand for the interpretation discussed in Bohm and Hiley \cite{bh93} and should be distinguished from what is called `Bohmian mechanics'.  Although both use exactly the same form of mathematics, the interpretations differ in some significant ways.}  (Bohm and Hiley \cite{bh87}, \cite{bh93} Holland \cite{h93}) does allow us to form a picture of such a process and reproduce all the known experimental results.  Indeed Bohm, Hiley and Dewdney \cite{bhd85} have already shown how the above Miller-Wheeler experiment can be understood in a consistent manner while maintaining the particle picture. There is no need to invoke non-locality here and the approach clearly shows there is no need to invoke the past only coming into being by action in the present.  

There exists a large volume of literature showing how the BI can also be used in many other typical quantum situations, allowing us to consistently account for these processes without the need for the type of explanation suggested by Wheeler.  Indeed application of the BI avoids some of the more spectacular paradoxes of SQM.  Particles do not go through both slits at the same time, cats do not end up in contradictory states such as being simultaneously alive and dead, and there is no measurement problem. 

In spite of all these results there is still a great reluctance to accept this explanation for reasons that we have never understood\footnote{John Polkinghorne \cite{p02} has recently written ``Certainly I have always felt the work you and Bohm did was far too readily dismissed by most physicists, who never gave it serious consideration and whose prejudice that it was `obviously wrong' was itself obviously wrong'' (Polkinghorne, private communication to BJH).}. This reluctance is shown by the many attempts to show the explanation is in some way wrong or predicts unacceptable features.  For example Englert, Scully, S\"{u}ssmann and Walther [ESSW] \cite{essw92}, Aharonov and Vaidman \cite{av96} and Aharonov, Englert and Scully \cite{aes99} have analysed the Bohm approach in detail and concluded that the explanation offered by the trajectories is too bizarre to be believable.  Unfortunately these analyses have not been carried out correctly and the conclusions they reach are wrong because they have not used the Bohm approach correctly.  The purpose of this paper is to clarify the Bohm approach as defined in Bohm and Hiley \cite{bh93} and to show what is wrong with the above arguments.  We will go on to show if treated correctly the Bohm approach does not produce the bizarre behaviour predicted by the above authors.  In fact the trajectories are essentially those that we would expect and there is no need for non-locality as previously suggested by D\"{u}rr, Fusseder and Goldstein \cite{dfg93} , Dewdney, Hardy and Squires \cite{dhs93} and indeed by ourselves in Hiley, Callaghan and Maroney \cite{hcm00}.

In this paper we will discuss these issues and correct the conclusion drawn by the various authors listed in the previous paragraph.  In particular the contents of this paper are as follows.  In section 2 we briefly re-examine the BI account presented in Bohm,  Hiley and Dewdney \cite{bhd85}  for the original delayed choice experiment outlined by Wheeler \cite{w78}.  In section 3 we move on to consider the delayed choice experiment introduced by ESSW \cite{essw92} where a microwave cavity is introduced into one of the arms of the interferometer.  This case is examined in detail both from the point of view of SQM and the BI.  We explain the principles involved by first replacing the cavity by a spin flip system.  This simplifies the mathematics so that we can bring out the principles involved more clearly. Then in section 4 we outline how the argument goes through with the cavity in place.  This requires us to use an extension of the BI applied to quantum field theory.  

We conclude that if the structure of the quantum potential that appears in the BI is correctly analysed we find there is no need for any non-local energy transfer in the above experiments.  In fact it is only the atom that goes through the cavity that gives up its excitation energy to the cavity even in the BI.  This is the opposite conclusion reached by Aharonov and Vaidman \cite{av96} and by ESSW.  Our results confirm that there is no difference between the account given by BI and that given by ESSW claiming to use SQM for these experiments.  Thus the conclusion that the Bohm trajectories are `metaphysical' or `surreal' does not follow from the arguments used by ESSW.  We finally discuss how our results provide the possibility of new insights into the role of measurement in quantum physics.


\section{Do quantum particles follow trajectories?}

There is a deeply held conviction as typified by Zeh \cite{z98} that a quantum particle cannot and does not have well-defined simultaneous values of position and momentum. This surely is what the uncertainty principle is telling us.  Actuality it is not telling us this.  What the uncertainty principle does say is that we cannot {\em measure} simultaneously the exact position and momentum of a particle.  This fact is not in dispute. But not being able to measure these values simultaneously {\em does not mean that they cannot exist simultaneously for the particle}. Equally we cannot be sure that a quantum particle actually {\em does not have simultaneous} values of these variables because there is no experimental way to rule out this possibility either.  The uncertainty principle only rules out simultaneous measurements. It says nothing about particles {\em having or not having} simultaneous $x$ and $p$.  Thus both views are {\em logically} possible. 

As we have seen Wheeler adopts an extreme position that not only do the trajectories not exist, but that the past does not exist independently of the present either. On the other hand the BI assumes particles could have exact values of position and momentum and then simply explores the consequences of this {\em assumption}.  Notice we are not {\em insisting} that the particles do actually have a simultaneous position and momentum.   How could we in view of the discussion in the previous paragraph?  

If we adopt the assumption that quantum particles do have simultaneous $x$ and $p$, which are, of course, unknown to us without measurement, then we must give up the insistence that the actual values of dynamical variables possessed by the particle are always given by the eigenvalues of the corresponding dynamical operators.  Such an insistence would clearly violate what is well established through theorems such as those of Gleason \cite{g57} and of Kochen and Specker \cite{ks67}.  All we insist on is that a measurement produces in the process of interaction with a measuring instrument an eigenvalue correspond to the operator associated with that particular instrument.  The particles have values for the complementary dynamical variable but these are not the eigenvalues of the corresponding dynamical operator {\em in the particular representation defined by the measuring instrument}.

This implies that the measurement can and does change the complementary variables.   In other words, measurement is not a passive process; it is an active process changing the system under investigation in a fundamental and irreducible way.  This leads to the idea that measurement is participatory in nature, remarkably a conclusion also proposed by Wheeler himself (see Patton and Wheeler \cite{pw75}). Bohm and Hiley \cite{bh93} explain in more detail how this participatory nature manifests itself in the BI.  It arises from the quantum potential induced by the measuring apparatus. We will bring this point out more clearly as we go along.

By assuming that a particle has simultaneously a precise position and momentum we can clearly still maintain the notion of a particle trajectory in a quantum process. Bohm and Hiley  \cite{bh93} and Holland \cite{h93} collect together a series of results that show how it is possible to define trajectories that are consistent with the Schr\"{o}dinger equation. The mathematics is unambiguous and exactly that used in the standard formalism. It is simply interpreted differently.  The equation for the trajectories contain, not only the classical potential in which the particle finds itself, but an additional term called the quantum potential, which suggests there is a novel quality of energy appearing only in the quantum domain. (See Feynman et al \cite{fls65}  for an alternative suggestion of this kind.) 

Both the trajectory and the quantum potential are determined by the real part of the Schr\"{o}dinger equation that results from polar decomposition of the wave function.  We find that the amplitude of the wave function completely determines the quantum potential.  In its simplest form this suggests that some additional physical field may be present, the properties of which are somehow encoded in the wave function.  One of the features of our early investigations of the BI was to find out precisely what properties this field must have in order to provide consistent interpretation of quantum processes.  It is the reasonableness of the physical nature of this potential and the trajectories that is the substance of the criticisms of Aharonov and Vaidman  \cite{av96} and of ESSW  \cite{essw92} and of Scully \cite{s98}.

To bring out the nature of these criticisms let us first recall how the original Wheeler interferometer can be discussed in the BI (see Bohm, Hiley and Dewdney  \cite{bhd85}).  As we have already indicated, the advantage of the BI in this case is that there exists a very straightforward explanation without the need of any suggestion of the type invoking ``present action determining the past".

We first consider the case where the incident particles are either electrons, or neutrons or even atoms and we restrict ourselves to the non-relativistic domain.  Here it is assumed that the particle {\em always} follows one and only one of the possible paths.  On the other hand the quantum field, described by $\psi $, satisfies the Schr\"{o}dinger equation which is defined globally.  The physical origins of this field will not concern us in this paper.  However our investigations suggest that this field is regarded as a field of {\em potentialities}, rather than a field of {\em actualities}.  This removes the situation that arises in SQM when we have two separated wave packets and then try to treat the packets as real.  It is this assumption of wave packets as being actual that leads to the debate about live and dead cats.  

In the BI, although there is talk of `empty wave packets' only one wave packet contains energy.  To conclude otherwise would violate energy conservation.  The `active' energy is the energy of the packet that contains the particle.  The empty packet remains a potentiality.  In an attempt to highlight this difference we are led to introduce the notions of `active information' and `passive information'.  The passive information cannot be discarded because it can become active again when the wave packets overlap later (see Bohm and Hiley \cite{bh93} for a detailed discussion of this point).  It should be noted that the arguments we use in the rest of this paper do not depend on this or any other specific interpretation of the quantum potential energy.  We have added these comments to enable the reader to relate to the discussion in Bohm and Hiley \cite{bh93}.


\section{Details of Wheeler's delayed choice experiment.}

Let us now turn to consider specific examples and begin by recalling the Wheeler delayed choice experiment using a two-beam interference device based on a Mach-Zender interferometer as shown in figure 1. We will assume the particles enter one at a time and each can be described by a Gaussian wave packet of width very much smaller than the dimensions of the apparatus so that the wave packets only overlap in regions $I_{1}$ and $I_{2}$.   Otherwise the wave packets have zero overlap.

The specific region of interest is $I_{2}$, which contains the movable beam splitter $BS_{2}$. In BI it is the quantum potential in this region that determines the ultimate behaviour of the particle.  This in turn depends upon whether the $BS_{2}$ is in place or not at the time the particle approaches the region $I_{2}$.  The position of $BS_{2}$ only affects the particle behaviour as it approaches the immediate neighbourhood of $I_{2}$.  Thus there is no possibility of the ``present action determining the past" in the way Wheeler suggests.  The past is past and cannot be affected by any activity in the present. This is because the quantum potential in $I_{2}$ depends on the actual position of $BS_{2}$ at the time the particle reaches $I_{2}$. We will now show how the results predicted by the BI agree exactly with the experimental predictions of SQM.

\subsection{Interferometer with $BS_{2}$ removed.}

Let us begin by first quickly recalling the SQM treatment of the delayed-choice experiment.  When $BS_{2}$ is removed (see figure 2) the wave function arriving at $D_{1}$ is  $ - \psi_{1}$.
 (The $ \frac{\pi}{2}$ phase changes arise from reflections at the mirror surfaces).
  This clearly gives a probability $|\psi_{1}|^{2}$ that $D_{1}$ fires.  The corresponding wave function arriving at $D_{2}$ is $i\psi_{2}$, giving a probability $|\psi_{2}|^{2}$ that $D_{2}$ fires. 
    
    If $BS_{1}$ is a 50/50 beam splitter, then each particle entering the interferometer will have a 50\% chance of firing one of the detectors.  This means that the device acts as a particle detector, because the particle will either take path 1, $BS_{1}M_{1}D_{1}$, trigging the detector $D_{1}$.  Or it will travel down path 2, $BS_{1}M_{2}D_{2}$, triggering detector $D_{2}$.
  \begin{figure}[htbp]
\begin{center}
\includegraphics[width=4in]{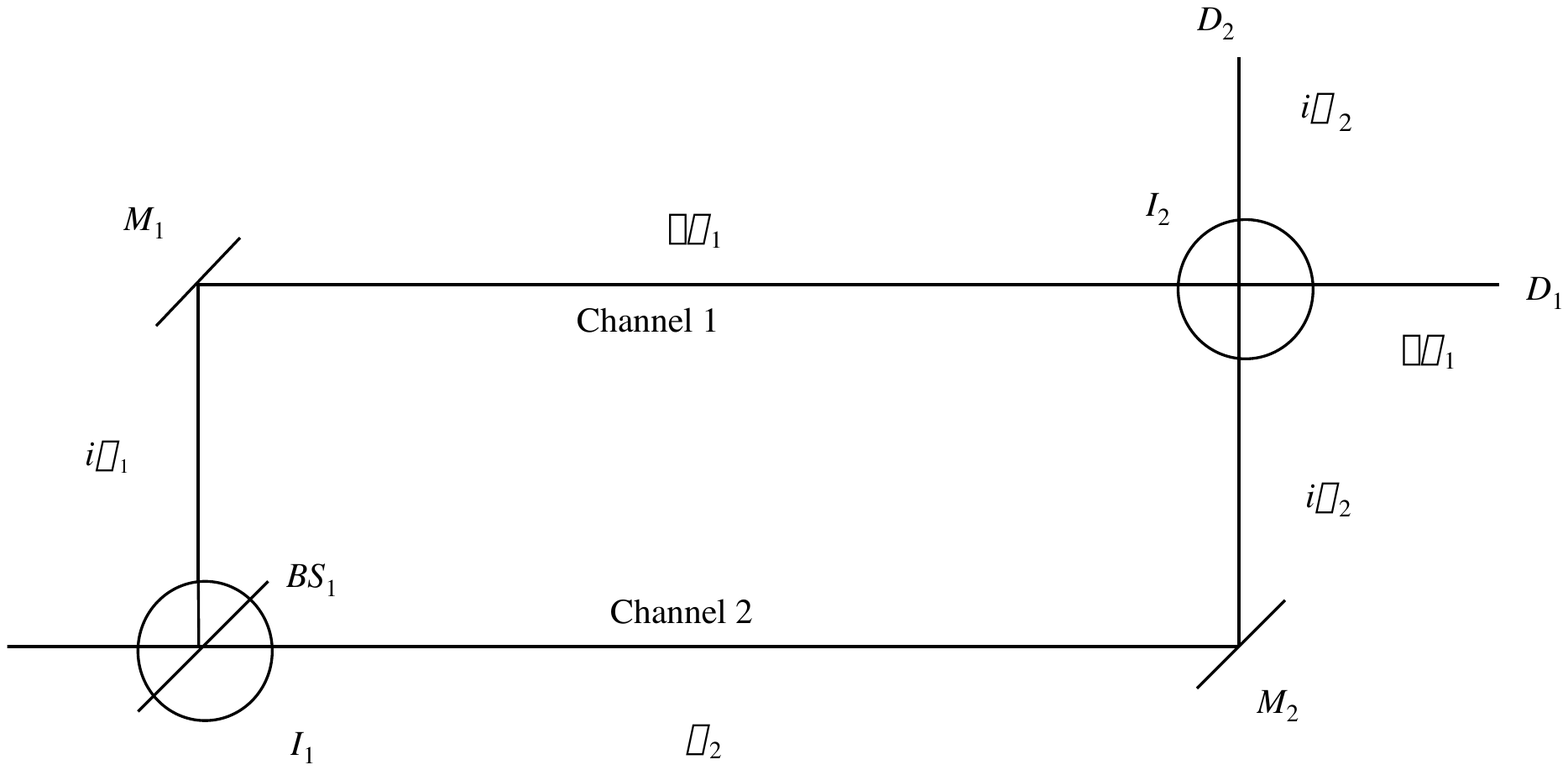}
\vspace{-3.8in}
\caption{ Interferometer acting as a particle detector.}
\end{center}
\end{figure}
Now let us turn to consider how the BI analyses this experiment.  Here we must construct an ensemble of trajectories, each individual trajectory corresponding to the possible initial values of position of the particle within the incident wave packet.  One set of trajectories will follow the upper arm of the apparatus, while the others follow the lower arm. We will call a distinct group of trajectories a `channel'. Thus the wave function in channel 1 will be  $\psi_{1}(\br,t) = R_{1}(\br,t)\exp[iS_{1}(\br,t)]  $ away from the regions $I_{1}$ and $I_{2}$ so that the Bohm momentum of the particle will be given by \begin{equation}
\bp_{1}(\br,t) = \nabla S_{1}(\br ,t)		
\end{equation}
 and the quantum potential acting on these particles will be given by
\begin{equation}
Q_{1}(\br ,t) = - \frac{1}{2m}\frac{\nabla^{2}R_{1}(\br ,t)}{R_{1}(\br ,t)}		
\end{equation}
There will be a corresponding expression for particles travelling in channel 2.

All of this is straightforward except in the region $I_{2}$, which is of particular interest to the analysis. Here the wave packets from each channel overlap and there will be a region of interference because the two wave packets are coherent. To find out how the trajectories behave in this region, we must write\footnote{For simplicity we will not write down normalised wave functions}
\begin{equation}
\Psi = -\psi_{1} +i \psi_{2} = R\exp[iS]			
\end{equation}
and then use
\begin{equation}
\bp = \nabla S \hspace{0.5cm}\mbox{and} \hspace{0.5cm}Q = - \frac{1}{2m}\frac{\nabla^{2}R}{R}		
\end{equation}
Thus to analyse the behaviour in the region $I_{2}$, we must write 
\[Re^{iS}=R_{1}e^{iS_{1}} + R_{2}e^{iS_{2}}  \]							
so that
\begin{equation}
R^{2} = R_{1}^{2} + R_{2}^{2} + 2R_{1}R_{2}\cos\Delta S'			
\end{equation}		 				
where $\Delta S' = S_{2} - S_{1}$.
Equation (5) clearly shows the presence of an interference term in the region $I_{2}$ since there is a contribution from each beam $\psi_{1}$ and $\psi_{2}$, which depends on the phase difference $\Delta S'$.  We show the behaviour of the quantum potential in this region in figure 4.

The particles following the trajectories then `bounce off' this potential as shown in figure 3 so that the particles in channel 1 end up triggering $D_{2}$, while the trajectories in channel 2 end up triggering $D_{1}$.  We sketch the overall behaviour of the channels in figure 5.  Notice that in all of this analysis the quantum potential is local.

 \begin{figure}[htbp]
\begin{center}
\includegraphics[width=3.2in]{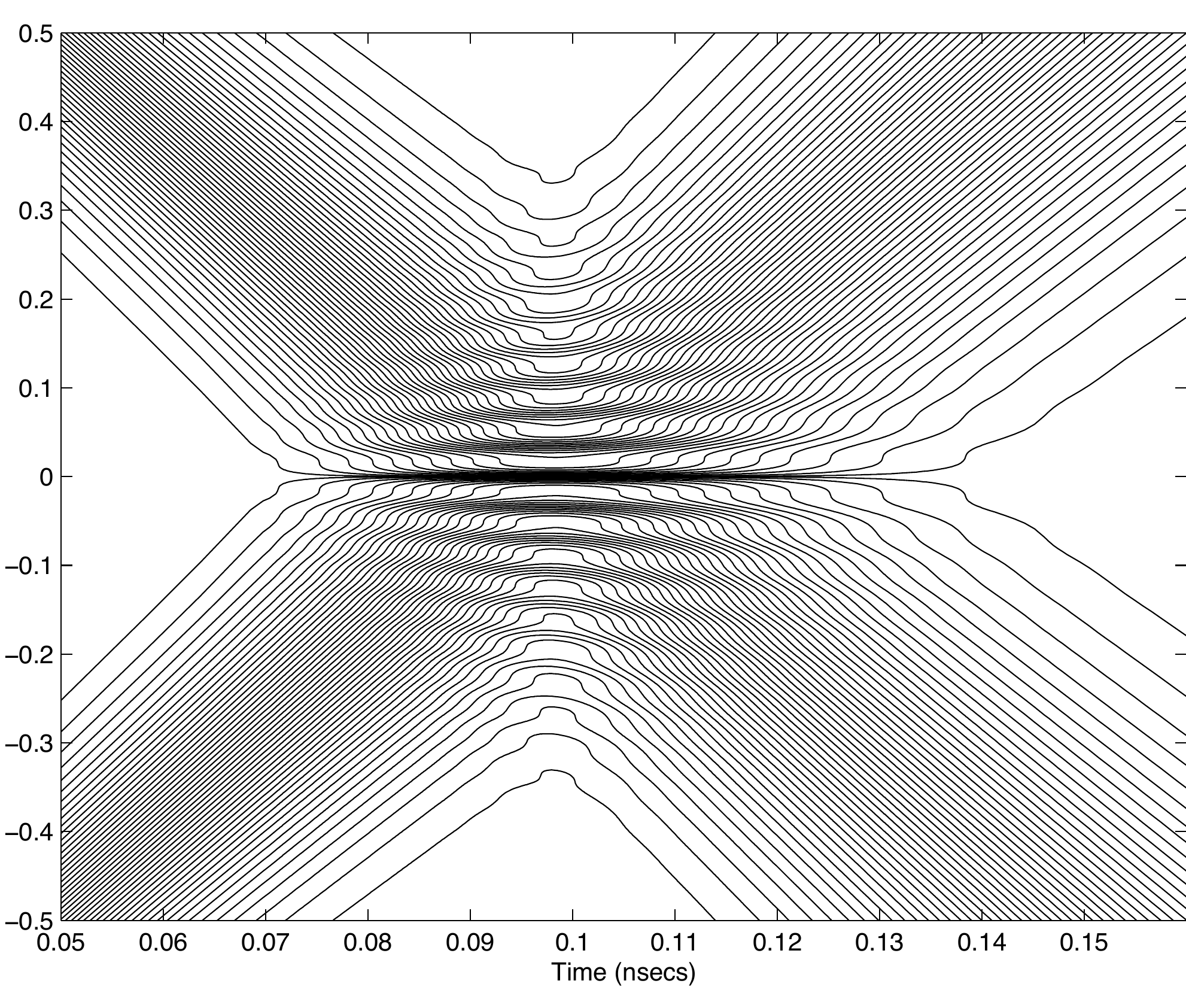}
\caption{Trajectories in the region $I_{2}$ without $BS_{2}$ in place.}
\label{default}
\end{center}
\end{figure}

\vspace{0.1in}

\begin{figure}[htbp]
\begin{center}
\includegraphics[width=3.5in]{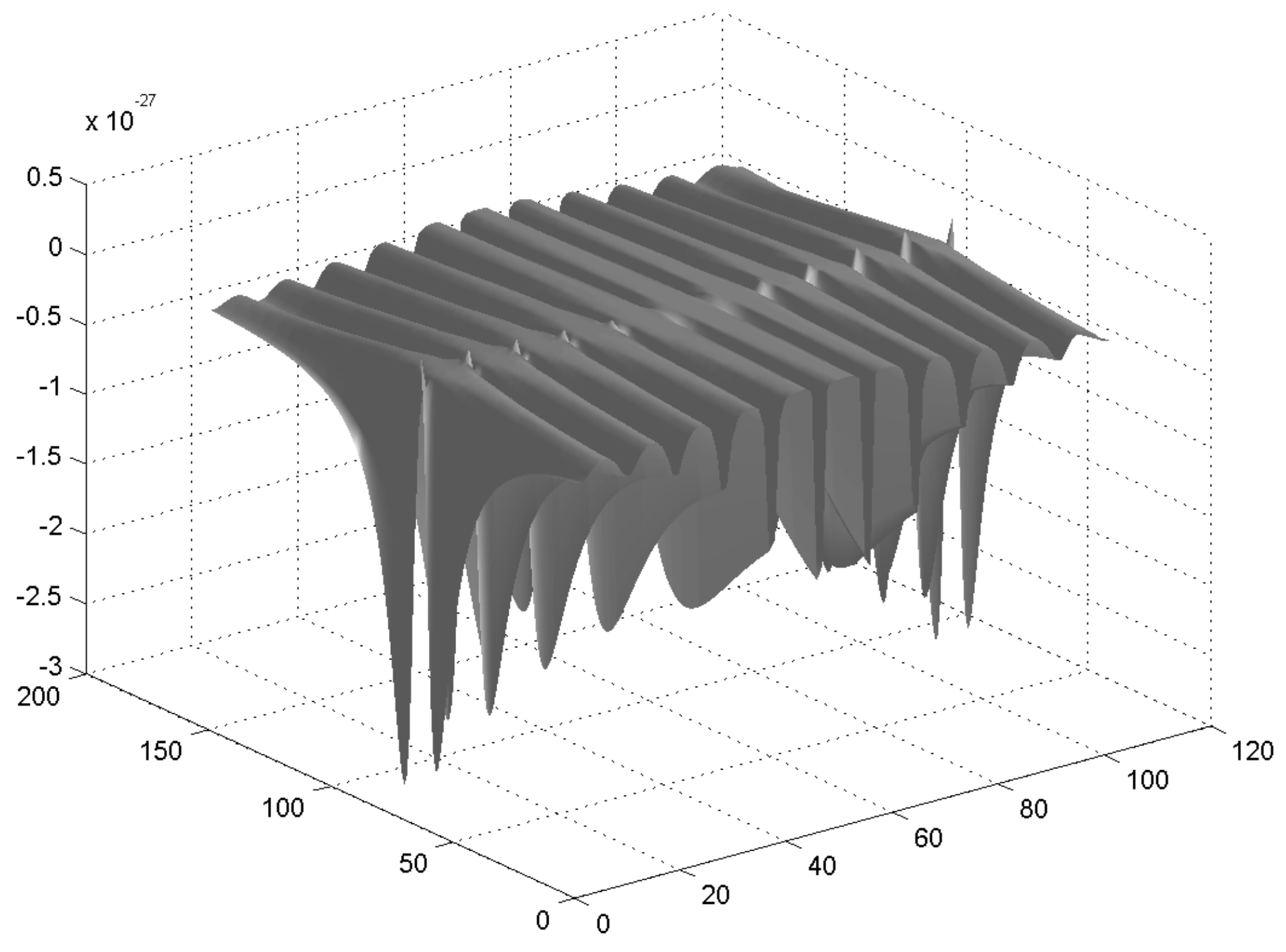}
\caption{Calculation of quantum potential in region $I_{2}$ without $BS_{2}$ in place.}
\end{center}
\end{figure}


\subsection{ Interference experiment with beam splitter $BS_{2}$ in place.}

Let us now consider the case when $BS_{2}$ is in place (see figure 6).  We will assume that beam splitter $BS_{2}$ is also a 50/50 splitter.  Using SQM the wave function at $D_{1}$ is 
\begin{equation}
\Psi_{D_{1}} = -(\psi_{1} +\psi_{2})			
\end{equation}
while the wave function at $D_{2}$ is
\begin{equation}
\Psi_{D_{2}} = i(\psi_{2}-\psi_{1})			
\end{equation}
Since $R_{1} = R_{2}$, and the wave functions are still in phase, the probability of triggering $D_{1}$ is unity, while the probability of triggering $D_{2}$ is zero.  This means that all the particles end up trigging $D_{1}$.  Thus we have 100\% response at $D_{1}$ and a zero response at $D_{2}$ and conclude that the apparatus acts as a wave detector so that we follow Wheeler \cite{w78} and say (loosely) that in SQM the particle ``travels down both arms", finally ending up in detector $D_{1}$.  In this case the other detector $D_{2}$ always remains silent. 
\begin{figure}[htbp]
\begin{center}
\includegraphics[width=4.5in]{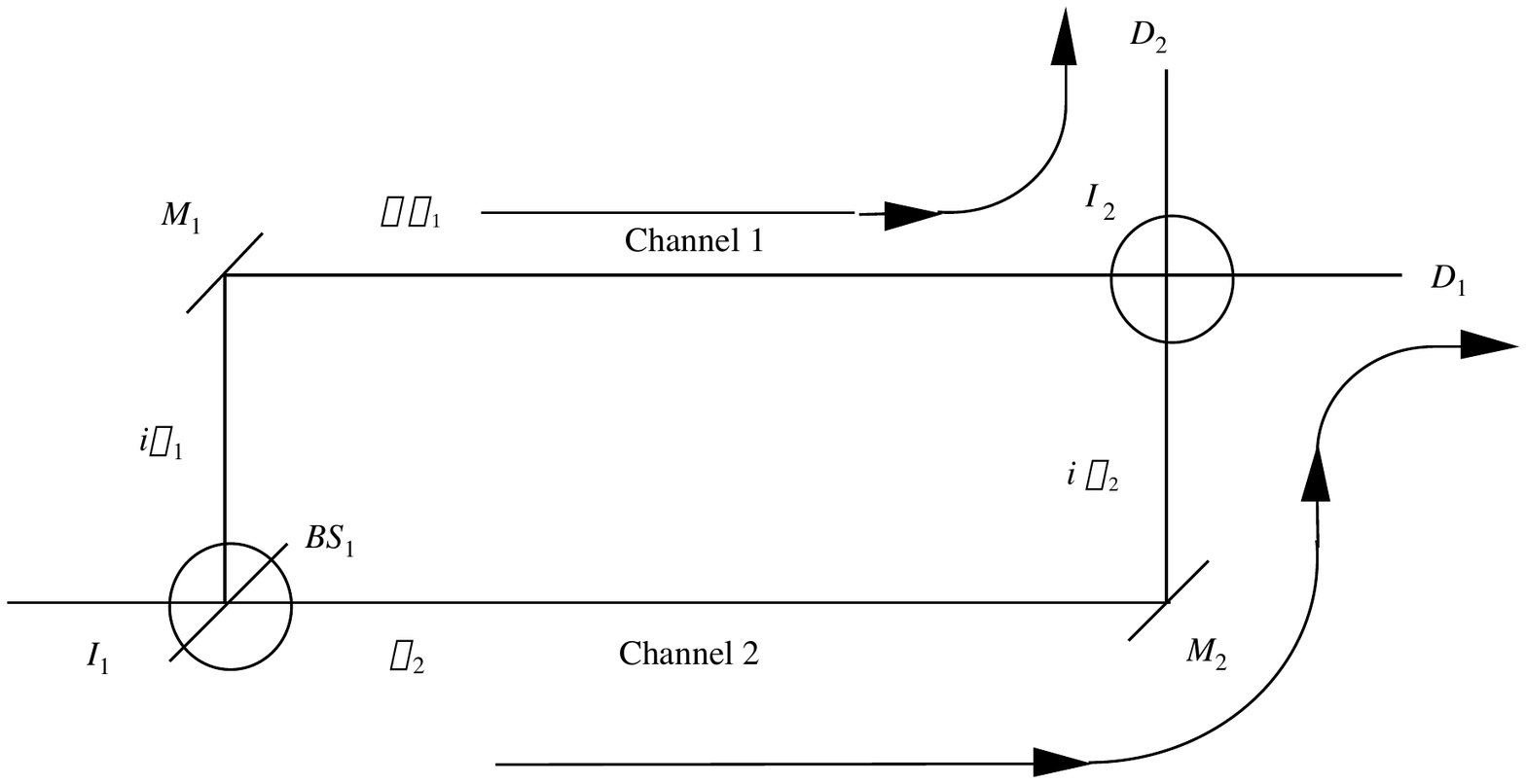}
\vspace{-3.9in}
\caption{Sketch of the Bohm trajectories without $BS_{2}$ in place.}
\end{center}
\end{figure}

How do we explain these results in the BI?  First we must notice that at beam splitter $BS_{1}$ the top half of the initial positions in the Gaussian packet are reflected while the bottom half are transmitted. This result is discussed in detail in Dewdney and Hiley \cite{dh82}.  As the two channels converge on beam splitter $BS_{2}$, the trajectories in channel 1 are now transmitted through it, while those in channel 2 are reflected. Thus all the trajectories end up trigging $D_{1}$.  It is straightforward to see the reason for this.  The probability of finding a particle reaching $D_{2}$ is zero and therefore all the particles in channel 1 must be transmitted.  The resulting trajectories are sketched in figure 7.
\begin{figure}[htbp]
\begin{center}
\includegraphics[width=4.5in]{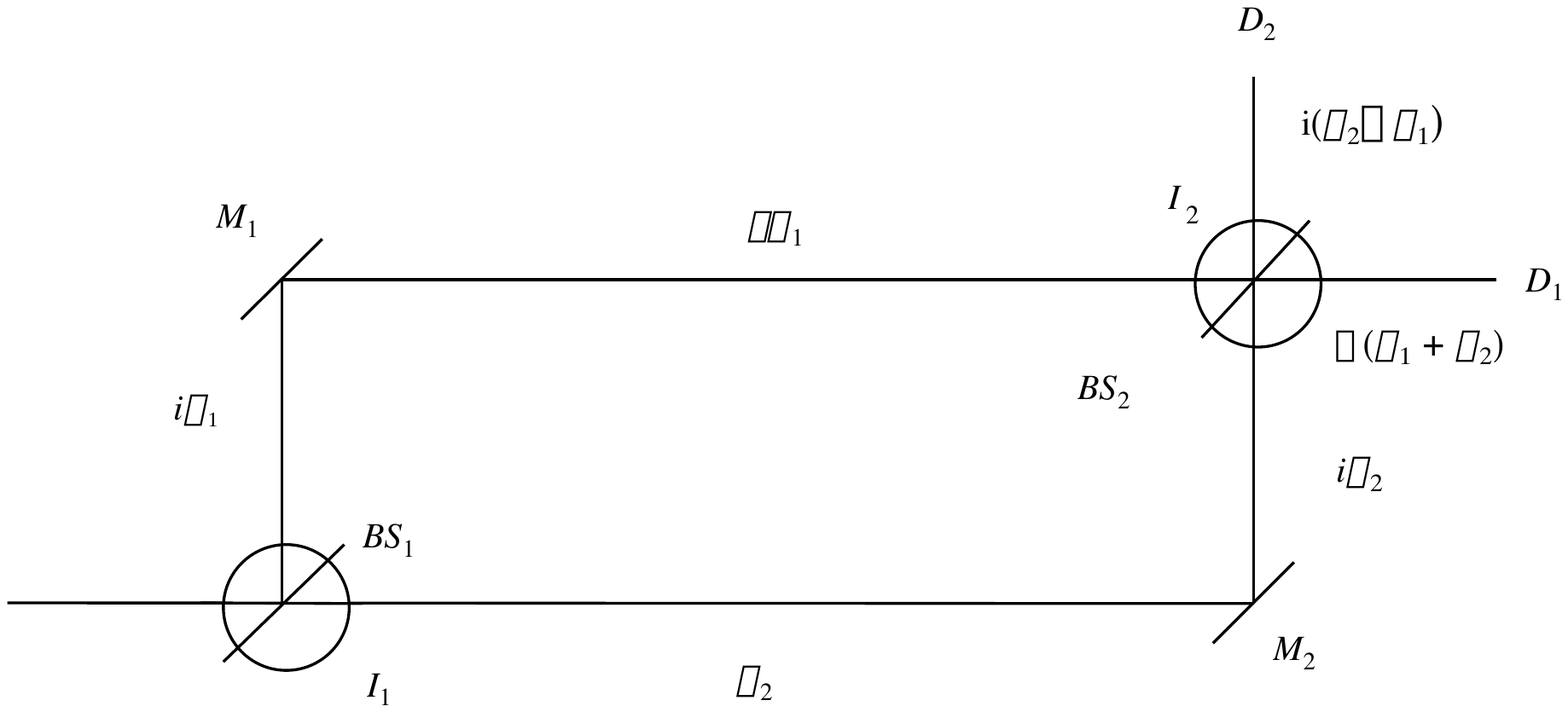}
\vspace{-4.3in}
\caption{Interferometer acting as a wave detector.}
\end{center}
\end{figure}

\subsection{ The delayed choice version of the interferometer.}

Now let us turn to consider what happens when beam splitter $BS_{2}$ can be inserted or removed once the particle has entered the interferometer, passing $BS_{1}$ but not yet reached $BS_{2}$.  We saw above that this caused a problem if we followed the line of argument used by Wheeler.  Applying the BI presents no such problem.  The particle travels in one of the channels regardless of whether $BS_{2}$ is in position or not.  The way it travels once it reaches the region $I_{2}$ depends on whether $BS_{2}$ is in position or not.  This in turn determines the quantum potential in that region, which in turn determines the ultimate behaviour of the particle.  

If the beam splitter is absent when a particle reaches $I_{2}$, it is reflected into a perpendicular direction no matter which channel it is actually in as shown in figure 5.  If $BS_{2}$ is in place then the quantum potential will be such as to allow the particle in channel 1 to travel through $BS_{2}$.  Whereas if the particle is in channel 2 it will be reflected at $BS_{2}$ so that all the particles enter detector $D_{1}$ as shown in figure 7. 
\begin{figure}[htbp]
\begin{center}
\includegraphics[width=4.5in]{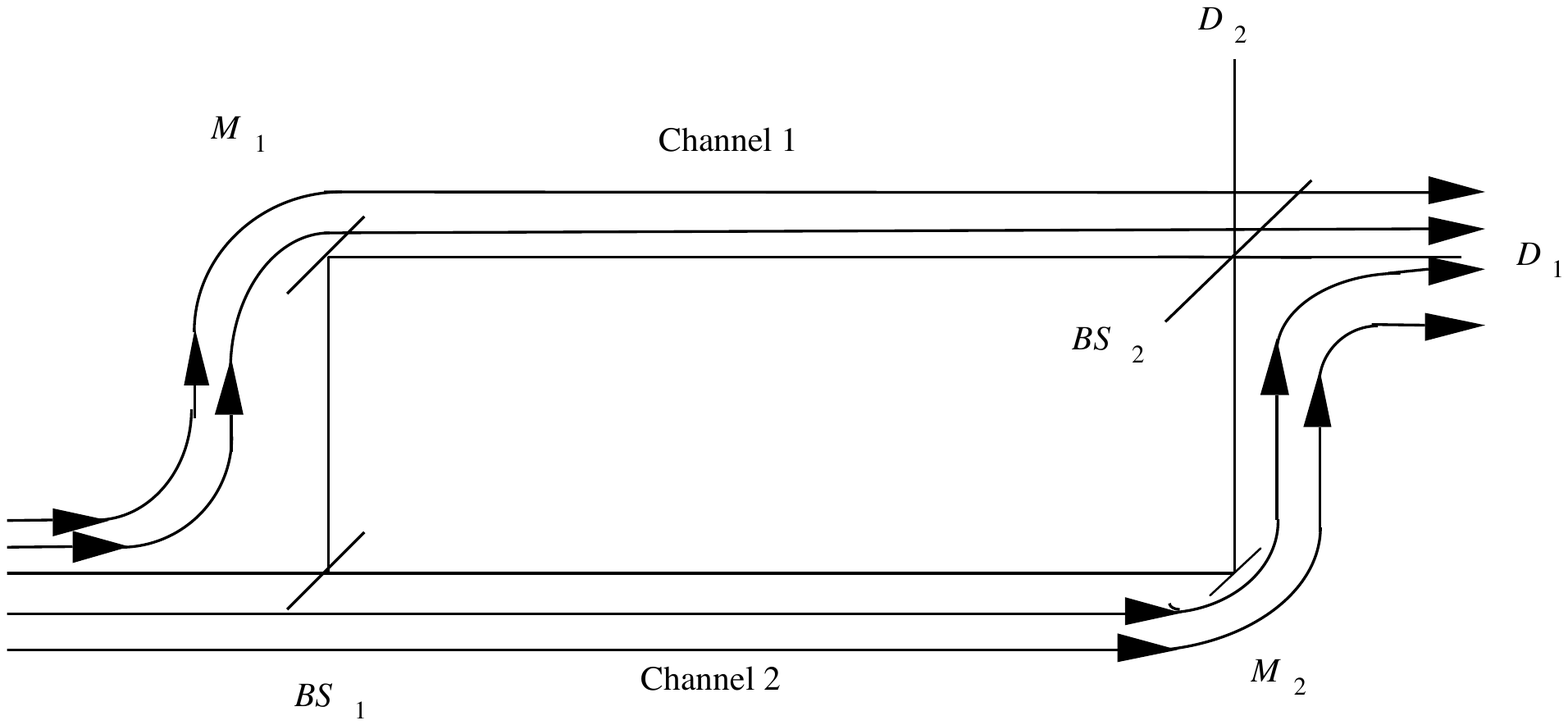}
\vspace{-4.3in}
\caption{Sketch of trajectories with $BS_{2}$ in place.}
\end{center}
\end{figure}

The explanation of the delayed choice results is thus very straightforward and depends only on the local properties of the quantum potential in the region of $I_{2}$ at the time the particle enters that region.  The value of the quantum potential in $I_{2}$ is determined only by the actual position of $BS_{2}$.  Hence there is no delayed choice problem here.  There is no need to claim that ``no phenomenon is a phenomenon unless it is an observed phenomenon". The result simply depends on whether $BS_{2}$ is in position or not at the time the particle reaches $I_{2}$ and this is independent of any observer being aware of the outcome of the experiment.  Remember BI is an ontological interpretation and the final outcome is independent of the observerÕs knowledge.

Note further that in these experiments the Bohm trajectories do not cross as correctly concluded by ESSW \cite{essw92}.  Let us now go on to see if this feature still holds in the modified experiments considered by ESSW.


\section{Variations of the Delayed-Choice Experiment.}

ESSW have modified the interferometer shown in figure 6 by removing $BS_{2}$ altogether and placing a micromaser cavity in one of the paths as shown in figure 8.
\begin{figure}[htbp]
\begin{center}
\includegraphics[width=4.5in]{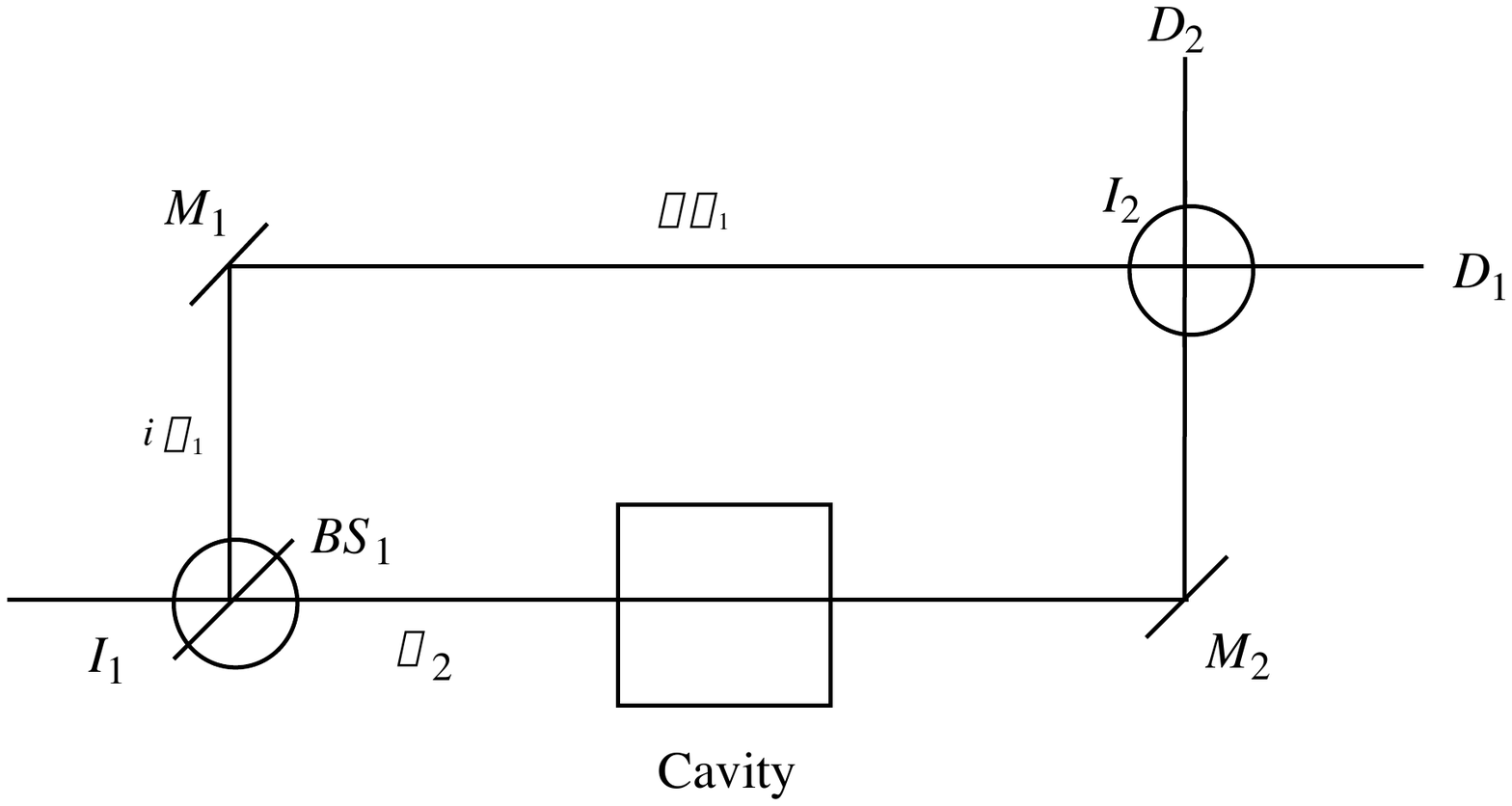}
\vspace{-4in}
\caption{ Interferometer with cavity in place and $BS_{2}$ removed.}
\label{default}
\end{center}
\end{figure}

This cavity has the key property that when a suitably excited atom passes through the cavity, it gives up all its internal energy of excitation to the cavity without introducing any random phase factors into the centre of mass wave function which continues unmodified.  This means that when the wave packet $\psi_{2} $ reaches $I_{2}$ it is still coherent with the wave packet $\psi_{1}$ that travels in channel 1.  Thus any loss of interference cannot be explained by the traditional assumption that it is loss of phase coherence that destroys interference.  This point has been clearly discussed and explained in Scully, Englert and Walther \cite{sew91}  and Scully and Walther \cite{sw98}and has been experimentally confirmed by D\"{u}rr, Nonn and Rempe  \cite{dnr98}.  We do not question the validity of this assumption.

What ESSW argued was that since there is still coherence in the region $I_{2}$, the behaviour of the Bohm trajectories should be as shown in figure 5.  Thus the particles travelling down channel 1 should trigger $D_{2}$, while those travelling down channel 2 should trigger $D_{1}$.  However, SQM and experiment show that the particles that trigger $D_{2}$ can lose their internal energy, whereas the particles triggering $D_{1}$ {\em never} lose any internal energy.  If the trajectories are as in figure 3 when the cavity is in place in channel 2 as ESSW maintain, then the particles that travel through the cavity never lose their energy while those not passing through the cavity can give up their internal energy to the cavity.  If this conclusion is correct then the Bohm trajectories are truly bizarre and surely ESSW would be right to conclude that these are not reliable and should be regarded as `surreal'.  But do the trajectories still behave as they are shown in figure 3?  

This conclusion was supported by a different experiment reported by Aharonov and Vaidman \cite{av96}.  They considered a bubble chamber rather than a microwave cavity and replaced the excited atoms with particles that can ionise the liquid molecules in the bubble chamber.  A significant part of their paper was concerned with an investigation into the relation between weak measurements and the BI.  This discussion is of no relevance to the discussions in this paper. However, they also considered what they called `robust' measurements and it is for these processes that they reach the same conclusion as ESSW.  They also conclude that the particle that does not pass through the bubble chamber causes the bubbles to appear.

Aharonov and Vaidman note that in their experiment the bubbles develop very slowly compared with the transition time of the particle.  The significance of this remark is that that the wave function of the apparatus `pointer' (i.e. the bubbles), although orthogonal in momentum space, do not significantly changed in position space because the bubbles take time to form.  So by the time they are formed to a significant radius, the particle has already reached the detector.  They claim that because of the slow speed of bubble formation, the trajectories are unaffected by the measurement and behave in exactly the same way as they would have behaved had no measurement been made, i.e., as shown in figure 3. 

Hence they conclude that slow development of the bubbles imply that `trajectories still don't cross', therefore the particles that do {\em not} go through the chamber somehow ionise the liquid. But Hiley, Callaghan and Maroney \cite{hcm00} have already shown that this conclusion is incorrect because Aharonov and Vaidman \cite{av96} had not used the BI correctly.  In fact Aharonov and Vaidman clearly state that they are not using the actual Bohm approach ``because the Bohm picture becomes very complex".  Indeed they make it very clear by writing
\begin{quote}
The fact that we see these difficulties follows from our particular approach to the Bohm theory in which the wave is not considered to be a reality.
\end{quote}
But the whole point of the BI is to assume the wave does have a `reality'.  This has been emphasised in all the key publications on the BI such as Bohm \cite{b52}, Bohm and Hiley \cite{bh93}, Holland \cite{h93} and D\"{u}rr, Goldstein and Zanghi \cite{dgz92}.  To emphasise the point further we will quote from Bohm and Hiley \cite{bh93}
\begin{quote}
As we have also suggested, however, this particle is never separated from a new type of quantum wave field that belongs to it and that it affects it fundamentally.
\end{quote}
In spite of the admission that they are using a different interpretation from BI, Aharonov and Vaidman \cite{av96}  go on to conclude that this unreasonable behaviour must also be attributed to the BI discussed in Bohm \cite{b52} and in Bohm and Hiley \cite{bh93} without giving any reasons for this conclusion.

This criticism was repeated again without further justification in Aharonov, Englert and Scully \cite{aes99}. We do not find their conclusion surprising.  Their model will always have this problem but, we emphasise again, this model is not the one proposed in Bohm and Hiley \cite{bh93}.  As we showed in Hiley, Callaghan and Maroney \cite{hcm00} (and will repeat the argument later in this section) in this case the trajectories {\em actually do cross}. Each particle that goes through the chamber ionises the liquid leaving a track upon which bubbles eventually form.  We will also show that one does not have to wait for the bubbles to form, it is sufficient to consider only the initial ionisation process from which a bubble will eventually form.  So there is a natural and reasonable local explanation of the whole process because it is the particles that do go through the bubble chamber that cause the ionisation and then move on to fire the detector $D_{2}$ even though the actual bubbles form later. 

The ESSW \cite{essw92} experiment in which the bubble chamber is replaced by a micromaser cavity requires a more subtle analysis, which was missed by Dewdney, Hardy and Squires \cite{dhs93}.  In this case there are no atoms to be ionised in the cavity, so it looks as if the final wave function describing the cavity will remain entangled with the wave function of the atom, giving what looks like a typical state that arises in the EPR situation.  Thus Dewdney, Hardy and Squires \cite{dhs93} argued that there must be a non-local exchange of energy between atom and cavity.  This seemed like a plausible explanation of the behaviour for those that are familiar with the BI.  Indeed it must be admitted that in our previous paper Hiley, Callaghan and Maroney \cite{hcm00} we came to a similar conclusion.  However here we show that this conclusion is wrong and that we do not need non-locality to account for the ESSW delayed choice experiment shown in figure 8.  All of this will be discussed in section 4.4.

 \subsection{The Aharonov-Vaidman version of the experiment.}

Let us then start with the Ahronov-Vaidman \cite{av96} version of the criticism of the BI because it is easier to bring out the error in their analysis. Firstly let us recall what happens according to SQM. In the region $I_{1}$ the wave function is $\Psi = \psi_{1} + \psi_{2}$.  If the particle triggers the bubble formation process we can regard the bubble chamber as acting like a measuring device and the particle gives up some of its energy causing the wave function to collapse to $\psi_{2}$.  As a consequence there is no interference in the region $I_{2}$ so that the particle goes straight through and fires detector $D_{2}$.  Thus the particles that trigger $D_{2}$ have lost some energy as can be checked experimentally.  On the other hand if the bubble formation process is not triggered, then the wave function collapses to $\psi_{1}$ so again the particle goes straight through the region $I_{2}$ with all its energy intact, eventually triggering $D_{1}$.  

What happens according to the BI?  There is no collapse so the wave function is still a linear combination with both $\psi_{1}$ and $\psi_{2}$ present in region $I_{2}$. At first sight it seems that if we use this wave function $\Psi$ to calculate the quantum potential in the region $I_{2}$ it should be the same as shown in figure 4.  This surely would mean  that the trajectories should be as shown in figure 3 mplying that the particles that travel in channel 1 eventually triggering $D_{2}$.  Since the particles that trigger $D_{2}$ can be shown to have lost some energy, it would mean that the BI predicts that bubble formation is triggered by the particle that does not go through the bubble chamber. 

The key question then is whether the coherence of the wave function  $\Psi = \psi_{1} + \psi_{2}$  is somehow `destroyed'.  One way out might be to appeal to the irreversibility involved in the bubble forming process.  However Aharonov and Vaidman \cite{av96} point out that the bubbles form relatively slowly so that they will not have formed until long after the particle has passed the region $I_{2}$.  This means that the effective wave function just after the particle has passed through $I_{2}$ is
\[ \Psi(\br ,y ,t)=[\psi_{1}(\br ,t)+\psi_{2}(\br ,t)]\Phi(y,t)\]
because as they put it ``Éthe density of the wave function is not changed significantly during the time of motion of the particle." (Aharonov and Vaidman \cite{av96a})  The implication here is that the apparatus wave function $\Phi(y,t)$  has not changed sufficiently before the particle arrives at the detector for us to write
\begin{equation}
\Psi (\br , y ,t)=\psi_{1}(\br ,t)\Phi_{NB}(y,t) + \psi_{2}(\br ,t)\Phi_{B}(y ,t)		
\end{equation}
where $\Phi_{NB}(y,t)[\Phi_{B}(y,t)]$   is the wave function of the bubble chamber when bubbles have not formed [have formed]. Wave function (8) could now be used in the standard Bohmian way to show that the quantum potential is no longer as shown in figure 4 (see Bohm and Hiley \cite{bh84} for details).

But all of this is irrelevant because measurement does not play a special role in the BI as it does in SQM. We must concentrate on the processes occurring at the particle level in the bubble chamber. Thus, when the particle enters the bubble chamber, the process that is central to the BI analysis is the ionisation process that takes place in the molecules of the liquid.  It is this ionisation that leads to a loss of coherence not because of irreversibility, but because the wave functions involved in the process no longer overlap and are spatially distinct. 

To show how this works we must first write down in detail the final wave function of all the particles involved in the ionisation process after ionisation has actually taken place. To make the argument as simple as possible and bring out clearly the principles involved, we will assume that the ionising particle that enters the bubble chamber will ionise one and only one liquid molecule and that, furthermore, there is 100\% chance of this happening.  We will sketch how to deal with a more realistic situation in section 4.3.1.

Let the wave function of the unionised liquid molecule be $\Psi_{UIL}(\br_{L},\br_{e})$   where $\br _{L}$ is the centre of mass co-ordinate of the liquid molecule and $\br_{e}$ is the position of the electron that will be ejected from the molecule on ionisation.  Immediately after the ionisation has taken place the wave function of the ionised molecule will be $\Psi_{IL}(\br_{L})$  and the wave function of the ejected electron will be $\phi(\br_{e})$.  The final wave function will then be
\begin{equation}
\Psi(\br ,\br_{L} ,\br_{e})=\psi_{1}(\br)\Psi_{UIL}(\br_{L} ,\br_{e})+\psi_{2}(\br)\Psi_{IL}(\br_{L})\phi(\br_{e})					
\end{equation}		
Here $\psi_{i}(\br)$, with $i = (1, 2)$, are the respective wave functions of the ionising particle at position $\br$. 

To work out what happens in the BI we must write the final wave function in the form
\begin{eqnarray} 
\Psi(\br ,\br_{L},\br_{e})  & = &  R(\br ,\br_{L},\br_{e})e^{iS(\br ,\br_{L}, \br_{e})}\nonumber  \\
& = & (R_{1}(\br)e^{iS_{1}(\br)}) (R_{UIL}(\br_{L} ,\br_{e})e^{iS_{UIL}(\br_{L} ,\br_{e})})\nonumber  \\
& + & (R_{2}(\br)e^{iS_{2}(\br)})(R_{IL}(\br_{L})e^{iS_{IL}(\br_{L})})(R_{e}(\br_{e})e^{iS_{e}(\br_{e})}) \nonumber
\end{eqnarray}
 And then use equation (5), which in this case becomes
\begin{equation}
R^{2} = (R_{1}R_{UIL})^{2} + (R_{2}R_{IL}R_{e})^{2} +2R_{1}R_{2}R_{UIL}R_{IL}R_{e}\cos\Delta S'
\end{equation}
We can calculate the quantum potential from this expression and see what effect this has in the region $I_{2}$.
 
Recall that the quantum potential must be evaluated for the actual positions of {\em all} the particles concerned.  Remember yet again that this is an ontological approach and the results do not depend on us {\em knowing} these positions.  The positions of all the particles are {\em actual} even though we do not know what these positions are.

The key to the disappearance of the interference term is the position of the ionised electron.  If the ionising particle passes along channel 1, there will be no ionisation so that the ionised electron will still be in the liquid molecule. Thus the probability of finding the electron outside the molecule is zero.  Hence $R_{e} = 0$ so the interference term in equation (10) will be zero, and therefore there will be no interference in region $I_{2}$.  This means that the atom will go straight to detector $D_{1}$.  

If however the ionising particle passes down channel 2 it will, by the assumption we are making about the efficiency of the ionisation process, ionise a liquid molecule.  In this case the probability of finding the electron in the unionised atom will be zero.  In this case $R_{UIL} = 0$ and again the interference in equation (10) vanishes.  As there is no interference in region $I_{2}$ the ionising particle goes straight through to trigger $D_{2}$.

If we look at this in terms of trajectories we find that because the interference term in equation (10) is always zero, trajectories now always cross.  The interference does not vanish because the ionising atom undergoes a randomisation of its phase, but because of final positions of particles involved in the interaction processes are such that their wave packets do not overlap and it is this fact that destroys interference.

Note the change in the position of the ionised electron is immediate and we do not have to wait until any bubbles form.  {\em The rate of bubble formation is irrelevant}.  The key point is that the ionised electron must be sufficiently far removed from the neighbourhood of the ionised liquid molecule so that the bubble can eventually form on the ionised molecule.  It is not necessary to invoke any principle of irreversibility at this stage.  All that is necessary is that the ionised electron does not return to the neighbourhood of the ionised liquid molecule before the bubble formation starts.  Of course irreversibility would help to ensure that this return would not take place at all but this is not essential here.  The essential point is to ensure {\em the probability of finding the electron back in its original molecule is zero so that the probability of the molecule remaining ionised is unity}.

In the above argument we have emphasised the role played by the quantum potential, but exactly the same result would be obtained if we use the guidance condition $\bp = \nabla S$  . We will show how all this works in more detail in a related example below (see equations (22)-(24)).

To summarise then we see that in this case the trajectories do cross.  Thus the BI trajectories do not behave in a `bizarre' fashion as claimed.  The ionising particles going through the bubble chamber, ionise the liquid and then go on to fire $D_{2}$.  While those that do not go through the bubble chamber, do not ionise the liquid and go straight on to fire $D_{1}$. There is no need to introduce any non-local exchange of energy. This is exactly what we would expect from quantum mechanics.  Therefore the conclusion drawn by Aharonov and Vaidman \cite{av96} and of Aharonov, Englert and Scully \cite{aes99} concerning the BI in this situation is simply wrong. 
   
\subsection{The ESSW experiment with the cavity in place.}

 Now let us move on to consider the subtler conditions introduced by ESSW \cite{essw92} and replace the bubble chamber with a micromaser cavity.  Here we have no ionised electron whose position has become changed by moving a significant distance so we cannot use the same argument. Something different must be involved in order to suppress the interference in region $I_{2}$ if we are not to get the bizarre results claimed.

We should also notice that we must change the argument even in the case of SQM because the cavity is not a measuring device in the same sense as a cloud chamber is a measuring device. So let us first remind ourselves how SQM deals with the interferometer with the cavity in place.  Let us write the state of the unexcited cavity as $|0\rangle_{c} $  while the excited cavity is written as $|1\rangle_{c} $ . The final wave function can be then written in mixed notation in the form
\begin{equation}
\Psi = \psi_{1}|0\rangle_{c} +\psi_{2}|1\rangle_{c} 		
\end{equation}
Here $\psi_{1}$ and $\psi_{2}$ describe the centre of mass wave function of the atom in channels 1 and 2 respectively.   Given the wave function (11), the probability of the final outcome is given by
\begin{equation}
|\Psi|^{2} = |\psi_{1}|^{2} + |\psi_{2}|^{2}				
\end{equation}
as the two cavity states are orthogonal. Thus there is a 50/50 chance of a particle triggering one of the detectors.  In fact because of the linearity of the Schršdinger equation, the wave packet in channel 1, $\psi_{1}$, will trigger $D_{1}$, while the wave packet in channel 2, $\psi_{2}$, will trigger $D_{2}$.  Indeed the probability of finding the cavity excited is given by
\begin{equation}
|_{c}\langle1|\Psi\rangle|^{2} = |\psi_{2}|^{2}			
\end{equation}
From this result it is reasonable to argue that the atom that travels in channel 2 gives up its internal energy to the cavity and then travels on to trigger $D_{2}$.  The atom that travels in channel 1, triggers $D_{1}$ and does not lose any internal energy because it does not go anywhere near the cavity.  From the standard point of view all of this is very satisfactory and very unremarkable.

\subsection{How does the Bohm Interpretation deal with this type of experiment?}

In order to bring out the principles involved in the BI we want to first replace the cavity, which would involve the mathematical complications of having to use quantum field theory, with a device that is simpler to deal with in the BI, but which presents the same conceptual problems as the cavity.  It is not that the BI cannot be applied to field theory (See for example Bohm, Hiley and Kaloyerou \cite{bhk87} and Bohm and Hiley \cite{bh93}) but that field theory brings in unnecessary complications that have little to do with the principles governing  the form of the quantum potential in the region $I_{2}$. Once the principles are clear  we can then return to discuss how to treat the behaviour of the quantum field in the cavity.

To this end recall the neutron scattering example discussed in Feynman \cite{f61}.  Consider a coherent beam of polarised neutrons being scattered off a polarised crystal target.  Here two processes are involved.  (1) The neutron scatters without spin flip or (2) the neutron produces a spin flip in the crystal. What Feynman argues is that there is no interference between the wave functions describing these two different processes even though the coherence between the neutron wave functions is maintained because, as he puts it, a spin flip is a potential measurement. 

Clearly the spin-flip example is different from the case of bubble chamber ionisation because the spin-flipped atom is heavy and it is assumed to have exactly the same position coordinate whether it has been flipped or not.  Thus we cannot use the spatial separation of wave packets to account for the lack of interference between the two channels when they meet again in the region $I_{2}$.  Something else must be involved.  

To see what this is let us consider the experiment where we replace the incident atoms by a polarised neutron beam and the cavity in figure 8 is replaced by a polarised crystal. Furthermore let us assume again for simplicity that the efficiency of the spin-flip process is 100\% so that whenever a neutron enters channel 2, a spin is flipped. We can make things conceptually even simpler by replacing the crystal by a single polarised atom.  We will also consider the idealised case that when one neutron passes the polarised atom there will be a spin-flip process every time, i.e. 100\% efficiency. This is not very realistic but it brings out the basic principles involved.   The wave function for this process will be
\begin{equation}
\Psi = \psi_{1}|\uparrow\rangle +\psi_{2}|\downarrow\rangle			
\end{equation}
where again $\psi$ is the centre of mass wave function of the neutron and the ket describes the spin state of the crystal atom.  Clearly since the spin states of the atom are orthogonal, the probability distribution the neutrons after they have passed through the region $I_{2}$ will have the same form as equation (11).  Since the probability of the detector $D_{1}$ firing will be given by
\begin{equation}
|\langle\downarrow|\Psi\rangle|^{2} = |\psi_{1}|^{2}					
\end{equation}
we can infer safely that the neutrons in channel 1 will pass straight through the region $I_{2}$ and trigger $D_{1}$.  Similarly those travelling in channel 2 trigger $D_{2}$.  Thus SQM analysis for this system is exactly the same as it is for the cavity.

Now let us turn to consider how the BI deals with this situation.  Recall that we must use wave functions throughout so that we must write equation (14) in the form
\begin{equation}
\Psi(\br_{1},\br_{2}) = \psi_{1}(\br_{1})\Phi_{\uparrow}(\br_{2}) + 			\psi_{2}(\br_{1})\Phi_{\downarrow}(\br_{2})						
\end{equation}
where $\Phi(\br_{2})$ is the wave function of the polarised scattering centre.  We must then write the final wave function in the form
\begin{equation}
\Psi =Re^{iS} = (R_{1}e^{iS_{1}})(R_{\uparrow}e^{iS_{\uparrow}}) +(R_{2}e^{iS_{2}})(R_{\downarrow}e^{iS_{\downarrow}})					
\end{equation}
From this we can determine the quantum potential which  in this case is given by
\begin{equation}														
Q = - \frac{1}{2m_{n}}\frac{\nabla_{\br_{1}}^{2}R}{R}  - \frac{1}{2m_{c}}\frac{\nabla_{\br_{2}}^{2}R}{R}
\end{equation}
If we assume that the crystal atom is heavy and has negligible recoil, only the first term in the quantum potential need concern us.  To evaluate this term we need to write $R$ in the form
\begin{equation}
R^{2} = (R_{1}R_{\uparrow})^{2} + (R_{2}R_{\downarrow})^{2} +2R_{1}R_{2}R_{\uparrow}R_{\downarrow}\cos\Delta S'					
\end{equation}
Now we come to the crucial point of our discussion. We need to evaluate this term for each {\em actual} trajectory.  Consider a neutron following a trajectory in channel 1.  Recall that the quantum potential is evaluated at the position of all the particles involved in the actual process.  This means we must take into account the contribution of the spin-state of the atom in channel 2.  But since we are assuming 100\% efficiency, the probability of finding the atom's spin flipped when the neutron is in channel 1 is {\em zero}.  This means that  $R_{\downarrow} = 0$  so that 
\begin{equation}
R^{2} = (R_{1}R_{\uparrow})^{2}				
\end{equation}
Thus there is no interference in the quantum potential in the region $I_{2}$ so the neutron goes straight through and triggers $D_{1}$.  Notice once again this is opposite to the conclusion reached by ESSW \cite{essw92} for the case of the cavity.

Now we will consider a neutron following a trajectory in channel 2.  Since in this case there is 100\% probability of finding the atom in a spin-flipped state and zero probability of finding the atom with spin up, we must now put $R_{\uparrow} = 0 $  .  This means that
\begin{equation}
R^{2} = (R_{1}R_{\downarrow})^{2}						
\end{equation}
So again there is no interference in the region $I_{2}$ and the neutrons go straight through to trigger $D_{2}$.

We can confirm this behaviour by looking directly at the phase and calculating the momentum of the neutron using
\begin{equation}
\bp_{n} = \nabla_{\br_{1}}S							
\end{equation}			
In the general case $S$ is given by 
\begin{equation}									
\tan S = \frac{(R_{1}R_{\uparrow})\sin (S_{1} + S_{\uparrow}) + (R_{2}R_{\downarrow})\sin (S_{2} +S_{\downarrow})}{(R_{1}R_{\uparrow})\cos (S_{1} + S_{\uparrow}) + (R_{2}R_{\downarrow})\cos (S_{2} + S_{\downarrow})}
\end{equation}
Thus for a neutron in channel 1 this reduces to 
\begin{equation}
S = S_{1}(\br_{1}) + S_{\uparrow}(\br_{2})			
\end{equation}
which confirms that the neutrons in channel 1 go straight through $I_{2}$ and trigger $D_{1}$.  Clearly for those neutrons in channel 2, $D_{2}$ is triggered.

Thus if the BI is analysed correctly we see that the behaviour of the trajectories is exactly the same as predicted by SQM using the arguments of ESSW \cite{essw92}.  So what has gone wrong with their argument in applying the BI?  The mistake they and others make stems from the behaviour of the trajectories shown in figure 3, in figure 7 and, incidentally, also from the trajectories calculated by  Philippidis, Dewdney and Hiley \cite{pdh79} for the two-slit interference experiment.  The characteristic feature of those trajectories is that they do not cross. By now it should be clear that we cannot assume in general that `trajectories do not cross'.  For example although we know the rule holds for systems described by wave functions of the form of equation (3), we know that it does not hold for {\em mixed states}. Here trajectories actually cross because there is no coherence between the separate components of the mixed state.  But it is not only in mixed states that they cross.  As we have shown above  they also cross for {\em pure states} of the type (8), even though the phases of the centre of mass wave functions are not randomised. Thus there is no universal rule for trajectories not crossing.  Each case must be considered separately for {\em all} types of pure states.

\subsubsection{What happens if the efficiency in less than 100\%?}

The example in the last sub-section assumed that the efficiency of the interaction was 100\%. This is actually very unrealistic so what happens in the case when the spin-flip is not 100\% efficient?  Suppose only a fraction $a^{2}$ of the neutrons induce a spin-flip.  Here the final wave function is
 \begin{equation}
\Psi = (\psi_{1} + b\psi_{2})|\uparrow\rangle +a\psi_{2}|\downarrow\rangle
\end{equation}
with $a^{2} + b^{2} = 1$  .  We now argue in the following way.  Divide the neutrons that travel in channel 2 into two groups, those that cause a spin-flip and those that do not.  If one of the neutrons cause a spin-flip, $\psi_{1}$ gives a zero contribution to the quantum potential by the argument given above so these neutrons travel straight through the region $I_{2}$ and trigger $D_{2}$.  Thus their behaviour is the same as the behaviour predicted by ESSW \cite{essw92}  using SQM.  

The rest see a quantum potential in the region $I_{2}$.  For this sub-set the quantum potential has an interference structure weakened by the factor b appearing in front of $\psi_{2}$.  These neutrons end up in detector $D_{1}$.  Thus while all the neutrons travelling in channel 2 that have not been involved in a spin-flip have their trajectories deflected to the detector $D_{1}$, the neutrons travelling in channel 1 can end up firing either $ D_{1}$ or $D_{2}$.  The fraction that will travel straight through $I_{2}$ will depend on the factor $b$.  Once again there is of course no problem with any non-local transfer of energy because these neutrons are not involved in any spin flip process.

\subsection{Treatment of the cavity in the Bohm Interpretation.}

To complete our description of the delayed choice experiments discussed above we must now show how the BI can be applied to the micromaser cavity.  Here we can no longer delay the argument and we must use field theory.  Fortunately the generalisation of the BI to enable it to be applied to field theory has already been discussed in Bohm \cite{b52}, Bohm, Hiley and Kaloyerou \cite{bhk87}, Bohm and Hiley \cite{bh93}, Kaloyerou \cite{k93} and Holland  \cite{h93}, \cite{h93a}.  We will not be concerned with all the details here but will be content to outline the principles, leaving the details for a later publication.

Let us return to consider figure 8 with the micromaser cavity in place.  We are assuming that when an excited atom enters the cavity, there is a local coupling between this excited atom and the field in the cavity described by a local interaction Hamiltonian given by
\begin{equation}
H_{I} = g\hat{\psi}_{a}(\br_{1})\hat{A}(\br_{1})\psi_{1}(\br_{1})														
\end{equation}
 
Here $\hat{\psi}_{a}(\br_{1})$   is the excited internal state of the atom, $\hat{A}_{1}(\br_{1})$  is the field variable in the cavity and  $\psi_{1}(\br_{1})$ is the centre of mass wave function which is not affected during the process. 

Standard quantum mechanics would write the wave function after the particles have passed through the Mach-Zender as
 \begin{equation}
\Psi(\br) = \psi_{1}(\br)|0\rangle_{c} + \psi_{2}(\br)|1\rangle_{c}          							  						 
\end{equation}
where, as before, $|0\rangle_{c}$  describes the state of the cavity when the atom does not pass through it and $|1\rangle_{c} $ is the excited state of the cavity after the atom has passed through it.  Since  $|0\rangle_{c}$ is orthogonal to $|1\rangle_{c} $, the intensity on the screen proportional to
\begin{equation}
|\Psi|^{2} =|\psi_{1}|^{2} + |\psi_{2}|^{2} 				
\end{equation}
which clearly shows no interference between the two beams.

How does the Bohm approach deal with this system?  First we need to write the wave function (27) in a form that is more appropriate for this approach, namely,
\begin{equation}										
\Psi(\br ,t) = \psi_{1}(\br)\Phi(\phi_{1}(\br_{1})) + \psi_{2}(\br)\Phi(\phi_{0}(\br_{1}))
\end{equation}										
Here $\Phi(\phi(\br_{1}))$ is the wave functional of the cavity field $\phi$, where $\phi_{1},(\phi_{0})$  is the excited (unexcited) field respectively.  Now we must write the final wave functional as
\begin{equation}
\Psi(\br,\Phi) = R(\br,\Phi)exp[iS(\br,\Phi)]						
\end{equation}
Then the trajectories can be evaluated from
\begin{equation}
\bp(\br,t) =\int d^{3}\br_{1}\nabla_{\br}S(\br,\Phi(\br_{1},t))		
\end{equation}											
while the quantum potential is now given by
\begin{eqnarray}										
Q = - \frac{1}{2}\int d^{3}\br_{1}\left [ \frac{1}{m}\frac{\nabla_{r}^{2}R}{R} + \frac { \frac{\delta^{2}R}{(\delta\Phi)^{2}}} {R} \right ]
\end{eqnarray}											

As has now become apparent, the use of fields has made the whole calculation more complicated.  However we again assume the effect of the interaction of the cavity on the centre of mass wave function of the atom is negligible so that we need only consider the contribution of the first term in equation (32) as we did in equation (18).  

To find the final effect of the cavity on the subsequent behaviour of the atom in region $I_{2}$ we must calculate the quantum potential effecting those atoms in channel 1 with the cavity in the unexcited state so the $R(É\phi_{e}(\br)É) = 0$.  Thus the atoms described by wave function $\psi_{1}$ are unaffected by $\psi_{2}$ so that they go straight through. Clearly those atoms in channel 2 also go straight through the region $I_{2}$.  Thus we draw exactly the same conclusion that we arrived at using the spin-flip argument.  The atom that goes through the cavity gives up its internal energy to the cavity and then goes straight through the region $I_{2}$ ending up triggering detector $D_{2}$.  This is exactly as predicted by standard quantum mechanics and there is no non-local exchange of energy between the atom arriving at $D_{2}$ since it actually passes through cavity, exchanging its energy locally.

Thus in all the cases we have considered in section 4 we get no bizarre features and we cannot conclude that the trajectories are `surreal'.  The behaviour is exactly as predicted by SQM using the arguments of ESSW \cite{essw92}.  Furthermore all energy exchanges are local.


\section{Measurement in quantum mechanics.} 

In the above analysis we have seen that the BI gives a perfectly acceptable account of how the energy is exchanged with the cavity and the claims that the trajectories are `surreal' have not been substantiated.  One of the confusions that seems to have led to this incorrect conclusion lies in role measurement plays in the BI.

One of the claims of the BI is that it does not have a measurement problem.  A measurement process is simply a special case of a quantum process.  One important feature that was considered by Bohm and Hiley \cite{bh84}  was to emphasise the role played by the macroscopic nature of the measuring instrument.  Their argument ran as follows.  During the interaction of this instrument with the physical system under investigation, the wave functions of all the components overlap and become `fused'.  This fusion process can produce a very complex quantum potential, which means that during the interaction the relevant variables of the observed system and the apparatus can undergo rapid fluctuation before settling into distinct sets of correlated movement.  For example, if the system is a particle, it will enter into one of a set of distinct channels, each channel corresponding to a unique value of the variable being measured. It should be noted that in this measurement process the complementary variables get changed in an unpredictable way so that the uncertainty principle is always satisfied, thus supporting the claim concerning participation made in section 2.

All of this becomes very clear if we consider the measurement of the spin components of a spin one-half particle using a Stern-Gerlach magnet.  As the particle enters the field of the Stern-Gerlach magnet, the interaction with the field produces two distinct channels.  One will be deflected `upwards' generating a channel that corresponds to spin `up'.  The other channel will be deflected `downwards' to give the channel corresponding to spin `down'. In this case there are no rapid fluctuations as the calculations of Dewdney, Holland and Kyprianidis \cite{dhk86} show.  Nevertheless the interaction with the magnetic field of the Stern-Gerlach magnet produces two distinct channels, one corresponding to the spin state, $\psi (+)$ and the other to $\psi (-)$.  There is no quantum potential linking the two beams as long as the channels are kept spatially separate.

Thus it appears from this argument that a necessary feature of a measurement process is that we must produce spatially separate and macroscopically distinct channels.  To put this in the familiar language of SQM, we must find a quantum process that produce separate, non-overlapping wave packets in space, each wave packet corresponding to a unique value of the variable being measured.  In technical terms this means that we must ensure that there is no intersection between the supports of each distinct wave packet, eg. $\sup(\psi_{i}) \cap \sup(\psi_{j}) = \O$  for $i \neq j$. 

Clearly this argument cannot work for the case of the cavity shown in figure 2.  Indeed $\sup(\phi_{0}) \cap \sup(\phi_{1}) \neq \O$   since both the excited and un-excited fields are supported in the same cavity.  This was one of the main factors why Dewdney, Hardy, and Squires, \cite{dhs93} and Hiley, Callaghan and Maroney \cite{hcm00} were content to introduce non-local exchanges of energy as a solution to the ESSW challenge. 

What these authors all assumed was that the non-overlapping of `wave packets' was a {\em necessary and sufficient} condition.  What we have argued above is that it is not a necessary condition but it is merely a {\em sufficient} condition.   What is necessary is for {\em there to be a unit probability of the cavity being in a particular energy eigenstate, all others, of course, being zero}.  What this ensures is that as long as the energy is fixed in the cavity, there will be no quantum potential coupling between the occupied particle channel and the unoccupied channel.  This also ensures that the particle will always behave in a way that is independent of all the other possibilities.  This means that in the example shown in figure 2 considered above, the atom passing through the cavity travels straight through the region $I_{2}$ and fires detector $D_{2}$.  This also means that in the language of SQM, the atom behaves as if the wave function had collapsed.  Again in this language it looks as if the cavity is behaving as a measuring instrument even though there has been no amplification to the macroscopic level and no irreversible process has occurred.  This is why Bohm and Hiley \cite{bh84}  emphasised that in the Bohm approach there is no fundamental difference between ordinary processes and what SQM chooses to call `measurement processes'.

Notice that {\em we do not need to know} whether the cavity has had its energy increased or not for the interference terms to be absent.  This is because we have an ontological theory, which means that there is a well-defined process actually occurring regardless of our knowledge of the details of this process. This process shows no interference effects in the region $I_{2}$ whether we choose to look at the cavity or not. We can go back at a later time after the particle has had time to pass through the cavity and through the region $I_{2}$ but before $D_{1}$ or $D_{2}$ have fired and find out the state of the cavity as ESSW \cite{essw92} have proposed. What we will find is this measurement in no way affects the subsequent behaviour of the atom and $D_{2}$ will always fire if the cavity is found in an excited state.  Thus there is no call for the notion such as ``present action determining the past" as Wheeler suggests. 

All we are doing in measuring the energy in the cavity at a later time is finding out which process {\em actually} took place.  The fact that this process may require irreversible amplification is of no relevance to the vanishing of the interference effects.  In other words there is no need to demand that the measurement is not complete until some irreversible macroscopic process has been recorded.  These results confirm the conclusions already established by Bohm and Hiley \cite{bh84} and there is certainly no need to argue ``no phenomenon is a phenomenon until it is an observed phenomenon."


\section{Conclusion.}

What we have shown in this paper is that some of the specific criticisms of the Bohm interpretation involving delayed choice experiments are not correct.  The properties of the trajectories that led Scully \cite{s98} to term them as `surreal' were based on the incorrect use of the BI.  Furthermore if the approach is used correctly then there is no need to invoke non-locality to explain the behaviour the particles in relation to the added cavity. The results then agree exactly with what Scully predicts using what he calls `standard quantum mechanics'. 

This of course does not mean that non-locality is removed from the BI.  In the situation discussed by Einstein, Podolsky and Rosen \cite{epr35}, it is the entangled wave function that produces the non-local quantum potential, which in turn is responsible for the corresponding non-local correlation of the particle trajectories.  The mistake that has been made by those attempting to answer the criticism is to assume wave functions of the type shown in equations (9), (16) and (29) are similar in this respect to EPR entangled states.  They are not. They are not because of the specific properties of systems like the micromaser cavity and polarised magnetic target.  The essential property of these particular systems is that we can attach a unit probability to one of the states even though we do not know which state this is.  The fact that a definite result has actually occurred is all that we need to know.  When this situation arises then all of the other potential states give no contribution to the quantum potential or the guidance condition so that there is no interference.  

This is not the same situation as in the case for the EPR entangled wave function.  In this case neither particle is in a well-defined individual state. This is reflected in the fact that there is only a 50\% chance of finding one of the two possible states of one of the particles.  Therefore the interference between the two states in the entanglement is not destroyed and it is this interference that leads to quantum non-locality.  However the interference is destroyed once we have a process that puts one of the particles into a definite state. In conventional terms this can be used as a record of the result and then the process is called a `measurement'.  But in the BI there is no need to record the result.  The fact that one result will occur with probability one is sufficient to destroy interference.  Thus delaying an examination of the `reading' is irrelevant.  The process has occurred and that is enough to destroy interference. 

It should be noted that in all of these discussions we offer no physical explanation of why there is a quantum potential or why the guidance condition takes the form it does.  The properties we have used follow directly from the Schr\"{o}dinger equation itself and the assumption that we have made about the particle possessing a simultaneous actual position and momentum.  As has been pointed out by Polkinghorne \cite{p02}, the key question is why we have the Schr\"{o}dinger equation in the first place.  Recently de Gosson \cite{g01} has shown that the Schr\"{o}dinger equation can be shown rigorously to exist in the covering groups of the symplectic group of classical physics and the quantum potential arises by projecting down onto the underlying group.  One of us, Hiley \cite{h03}, \cite{h04} has recently argued that a similar structure can arise by regarding quantum mechanics as arising from a non-commutative geometry where it is only possible to generate manifolds by projection into the so-called `shadow' manifolds.  Here the mathematical structure is certainly becoming clearer but the implications for the resulting structure of physical processes need further investigation.


\begin{thebibliography}{99}

\bibitem{w78}Wheeler, J. A., [1978], The ÒPastÓ and the ÒDelayed-ChoiceÓ Double-slit Experiment, in {\em Mathematical Foundations of Quantum Theory}, ed. Marlow, pp. 9-47, Academic Press, New York.
\bibitem{mw83}Miller, W. A. and Wheeler, J. A., [1983], Delayed-Choice Experiments and Bohr's Elementary Quantum Phenomenon, p.140-51.  {\em Proc. Int. Symp. Found. of Quantum Mechs}. Tokyo.  (Physical Society of Japan)
\bibitem{gz97}Greenstein, G and Zajonc, A. G., [1997], {\em The Quantum Challenge}, Jones and Bartlett, Sudbury MA.
\bibitem{b61}Bohr, N., [1961], {\em Atomic Physics and Human Knowledge}, p. 51, Science Editions, New York. 
\bibitem{bh87} Bohm, D. and Hiley, B. J., [1987], An Ontological Basis for Quantum Theory: I - Non-relativistic Particle Systems, {\em Phys. Reports} {\bf 144}, 323-348. 
\bibitem{bh93} Bohm, D. and Hiley, B. J., [1993], {\em The Undivided Universe: an Ontological Interpretation of Quantum Theory}, Routledge, London.
\bibitem{h93}Holland, P. R., [1993], {\em The Quantum Theory of Motion}, Cambridge University Press, Cambridge.
\bibitem{bhd85} Bohm D., Hiley, B. J., and Dewdney, C., [1985], A Quantum Potential Approach to the Wheeler Delayed-Choice Experiment.  {\em Nature}, {\bf 315}, 294-297.
\bibitem{essw92}Englert, J., Scully, M. O., S\"{u}ssman, G. and Walther, H., [1992], Surrealistic Bohm Trajectories, {\em Z. Naturforsch}. {\bf 47a}, 1175-1186.
\bibitem{av96} Aharonov, Y. and Vaidman, L. [1996]  About Position Measurements which do not show the Bohmian Particle Position, in {\em Bohmian Mechanics and Quantum Theory: an Appraisal}, ed J. T. Cushing, A Fine and S. Goldstein, Boston Studies in the Philosophy of Science, 184, 141-154, Kluwer, Dordrecht.
\bibitem{av96a} Aharonov, Y. and Vaidman, L. [1996], {\em ibid}, p. 151.
\bibitem{aes99} Aharonov, Y., Englert, B-G. and Scully, M. O., [1999], Protective measurements and Bohm trajectories, {\em Phys. Lett.}, {\bf A263}, 137-146.
\bibitem{dfg93}D\"{u}rr, D., Fusseder,W and Goldstein, S. [1993], Comment on ``Surrealistic Bohm Trajectories'', {\em Z. Naturforsch}. {\bf 48a}, 1261-2.
\bibitem{dhs93}Dewdney, C., Hardy, L. and Squires, E. J., [1993], How late Measurements of Quantum Trajectories can fool a Detector, {\em Phys. Lett.}, {\bf 184}, 6-11.
\bibitem{hcm00}Hiley, B. J., Callaghan, R. E.  and Maroney, O., [2000], Quantum Trajectories, Real, Surreal, or an Approximation to a Deeper Process. {\em quant-ph/0010020}.
\bibitem{p02}Polkinghorne, J., [2002], {\em Quantum Theory: a very short introduction}, Oxford University Press, Oxford.
\bibitem{z98} Zeh, D. [1998], Why BohmÕs Quantum Theory? {\em Quant-ph/9812059}.
\bibitem{g57}Gleason, A. M., [1957], Measures on the Closed Sub-spaces of Hilbert Space, {\em J. Maths. Mechs.}, {\bf 6}, 885-93.
\bibitem{ks67}Kochen, S. and Specker, E. P., [1967], The Problem of Hidden Variables in Quantum Mechanics, {\em J. Math. Mech}.  {\bf 17}, 59-87.
\bibitem{pw75}Patton, C. M., and Wheeler, J. A., [1975], Is physics legislated by cosmogony? {\em In Quantum Gravity}, eds. Isham, C., Penrose, R., and Schama, D., pp. 538-605, Clarendon Press, Oxford.
\bibitem{fls65}Feynman, R. P., Leighton, R. B. and Sands, M., [1965], {\em The Feynman Lectures on Physics} III, p. 21-12, Addison-Wesley, Reading, Mass.
\bibitem{s98}Scully, M. O., [1998], Do Bohm trajectories always provide a trustworthy physical picture of particle motion, {\em Phys. Scripta}, {\bf T76}, 41-46.
\bibitem{dh82} Dewdney, C. and Hiley, B. J.,[1982],   A Quantum Potential Description of One-dimensional Time-dependent  Scattering from Square Barriers,  {\em Found. Phys}. {\bf 12}, 27-48.
\bibitem{sew91}  Scully, O. M., Englert, B. G. and Walter, H., [1991], Quantum Optical Tests of Complementary, {\em Nature}, {\bf 351}, 111-116.
\bibitem{sw98}Scully, O. M., and Walther, H., [1998], An Operational Analysis of Quantum Eraser and Delayed Choice, {\em Found. Phys}., {\bf 28}, 399-413.
\bibitem{dnr98}D\"{u}rr, S., Nonn, T. and Rempe, G., [1998], Origin of quantum mechanical complementarity probed by a Ôwhich-wayÕ experiment in an atom interferometer, {\em Nature}, {\bf 395}, 33-37.
\bibitem{b52}Bohm, D., [1952], A Suggested Interpretation of the Quantum Theory in Terms of Hidden Variables, I, {\em Phys. Rev.}, {\bf 85}, and II, 166-179; {\bf 85},180-193.
\bibitem{dgz92}D\"{u}rr, D., Goldstein, S. and Zanghi, N., [1992], Quantum Equilibrium and the Origin of Absolute Uncertainty, {\em J. Stat. Phys.},  {\bf 67},  843-907.
\bibitem{bh84} Bohm, D. and Hiley, B. J., [1984], Measurement Understood Through the Quantum Potential Approach, {\em Found. Phys}., {\bf 14}, 255-74.
\bibitem{bhk87} Bohm, D. and Hiley, B. J. and Kaloyerou, P.N., [1987], An Ontological Basis for the Quantum Theory: II -A Causal Interpretation of Quantum Fields, {\em Phys. Reports}, {\bf 144}, 349-375. 
\bibitem{f61}Feynman, R. P., [1961], {\em Theory of Fundamental Processes}, p. 3, Benjamin, New York.
\bibitem{pdh79}Philippidis, C., Dewdney, C. and Hiley, B. J., [1979], Quantum Interference and the Quantum Potential, {\em Nuovo Cimento}, {\bf 52B},  15-28.
\bibitem{bhk87} Bohm, D., Hiley, B. J. and Kaloyerou, P.N., [1987] An Ontological Basis for the Quantum Theory: II -A Causal Interpretation of Quantum Fields,   {\em Physics Reports}, {\bf 144}, 349-375. 
\bibitem{k93}Kaloyerou, P. N., [1993], The Causal Interpretation of the Electromagnetic Field, {\em Physics Reports}. {\bf 244}, 287-385.
\bibitem{h93a}Holland, P. R., [1993a], The de Broglie-Bohm Theory of motion and Quantum Field Theory, {\em Physics Reports}, {\bf 224}, 95-150.
\bibitem{dhk86}Dewdney, C., Holland, P. R. and Kyprianidis, A.,[1986], {\em Phys. Letts.}, {\bf 119A}, 259.
\bibitem{epr35}Einstein, A., Podolsky, B., and Rosen, N., [1935], Can Quantum-Mechanical Description of Physical Reality be Considered Complete, {\em Phys. Rev.}, {\bf 47}, 777-80.
\bibitem{g01}de Gosson, M., [2001], {\em The Principles of Newtonian and Quantum Mechanics: the need for Plank's Constant}, Imperial College Press, London, 2001.
\bibitem{h03}Hiley, B. J., [2003], Phase Space Descriptions of Quantum Phenomena, {\em Proc. Int. Conf. Quantum Theory: Reconsideration of Foundations 2}, V\"{a}xj\"{o}, Sweden.
\bibitem{h04}Hiley, B. J., [2005], Non-commutative Quantum Geometry: a re-appraisal of the Bohm approach to quantum theory, in {\em Quo Vadis Quantum Mechanics}, ed A. Elitzur, S. Dolev, and N. Kolenda, pp. 299-324,Springer, Berlin.

\end{thebibliography}
 \end{document}